\documentclass[aps,prl,twocolumn,groupedaddress,showpacs,floats]{revtex4}%

\usepackage{graphicx}
\usepackage{amssymb}
\usepackage[dvipdfm]{hyperref}
\usepackage{amsmath}

\begin{document}

\title{Quantitative Determination of Eliashberg Function and Evidence of Strong
Electron Coupling with Multiple Phonon Modes in Heavily Over-doped
(Bi,Pb)$_2$Sr$_2$CuO$_{6+\delta}$}

\author{Lin Zhao$^{1}$, Jing Wang$^{2}$}

\author{Junren Shi$^{2}$}
\email[Corresponding author: ]{jrshi@aphy.iphy.ac.cn}

\author{Wentao Zhang$^{1}$, Haiyun Liu$^{1}$, Jianqiao
Meng$^{1}$, Guodong Liu$^{1}$, Xiaoli Dong$^{1}$, Wei Lu$^{1}$,
Guiling Wang$^3$, Yong Zhu$^3$, Xiaoyang Wang$^{3}$, Qinjun
Peng$^{3}$, Zhimin Wang$^{3}$, Shenjin Zhang$^{3}$, Feng Yang$^{3}$,
Chuangtian Chen$^{3}$, Zuyan Xu$^{3}$}

\author{X. J. Zhou$^{1,}$}
\email[Corresponding author: ]{XJZhou@aphy.iphy.ac.cn}

\affiliation {
\\$^1$National Lab for Superconductivity, Beijing National Laboratory
for Condensed Matter Physics,Institute of Physics, Chinese Academy
of Science, Beijing 100190, China
\\$^2$Beijing National Laboratory for Condensed
Matter Physics, Institute of Physics, Chinese Academy of Science,
Beijing 100190, China
\\$^3$Technical Institute of Physics and Chemistry, Chinese
Academy of Sciences, Beijing 100190, China }

\date{January 31, 2010}
%
%

\begin{abstract}

Super-high resolution laser-based angle-resolved photoemission
spectroscopy measurements have been carried out on a heavily
overdoped (Bi,Pb)$_2$Sr$_2$CuO$_{6+\delta}$ ($T_c\sim 5$~K)
superconductor. Taking advantage of the high-precision data on the
subtle change of the quasi-particle dispersion at different
temperatures, we develop a general procedure to determine the bare
band dispersion and extract the bosonic spectral function
quantitatively. Our results show unambiguously that the $\sim$70 meV
nodal kink is due to the electron coupling with the multiple phonon
modes, with a large mass enhancement factor  $\lambda \sim 0.42$
even in the heavily over-doped regime.
\end{abstract}


\maketitle

The interaction of electrons with phonons and other collective
excitations (bosons) in solids dictates fundamental physical
properties of materials~\cite{GGrimvall}.  In a Fermi liquid
picture, such interaction can be described by the Eliashberg
spectral function $\alpha^{2}F(\omega)$, which accounts for the
bosonic modes involved in the coupling and their strengths. In the
conventional superconductors, extraction of the Eliashberg function
played a key role in pinning down the phonon as the glue for the
electron pairing that gives rise to the BCS
superconductivity~\cite{JMRowell}. For high temperature cuprate
superconductors and other complex compounds, one expects that the
extraction of the Eliashberg function will also be important in
understanding the exotic physical properties and the mechanism of
high temperature superconductivity~\cite
{Tunneling,Optical,ADamascelli,JunrenMEM,XJZhouMEM,NonBi2201}.

The electron-boson coupling gives rise to the band renormalization
and the change of the quasiparticle scattering rate, which can be
described by the real part (Re$\Sigma(k, \omega)$) and the imaginary
part (Im$\Sigma(k, \omega)$) of the electron self-energy $\Sigma(k,
\omega)$, respectively. With the dramatic improvement of resolution,
angle-resolved photoemission spectroscopy (ARPES) has emerged as a
powerful tool in probing such many-body effects~\cite{ADamascelli}.
Under the sudden approximation, it measures the single particle
spectral function
\begin{math}
A(k,\omega) =
({1}/{\pi}){\mathrm{Im}\Sigma(k,\omega)}/\{[\omega-\epsilon^0_{k}-\mathrm
{Re}\Sigma(k,\omega)]^{2}+[\mathrm{Im}\Sigma(k,\omega)]^{2}\}
\end{math}, which is directly related to the electron self-energy~\cite{ADamascelli}.  ARPES
has been extensively employed in investigating many-body effects in
conventional metals~\cite{BeSystem,TValla},
graphene~\cite{EliGraphene} and complex
materials~\cite{ADamascelli,Manganites}.  However, a long-standing
issue with the ARPES technique is the ambiguous ``bare band"
$\epsilon^0_{k}$\cite{DresdenTry}, which hinders a quantitative
determination of the electron self-energy
$\mathrm{Re}\Sigma$(k,$\epsilon$)=$\epsilon$-$\epsilon^0_{k}~$ and
the underlying bosonic spectral function $\alpha^2
F(\omega)$~\cite{JunrenMEM,XJZhouMEM,NonBi2201}.

Coming specifically to high-$T_c$ cuprate superconductors, although
there is a general consensus on the electron-boson coupling as the
origin of the universally observed 50$\sim$80 meV dispersion kink
along the (0,0)-($\pi$,$\pi$) nodal
direction~\cite{PVBogdanov,ALanzara,PJohnson,AKaminski,XJZhouNature,AAKordyuk,WTZhang1},
it remains under debate on the nature of the boson(s) involved,
mainly between phonons~\cite{ALanzara} and magnetic
origins~\cite{AAKordyuk,ARPESNeutron,Takahashi}. Theoretical
calculations suggest that the electron-phonon coupling in cuprates
is too weak to account for the nodal kink~\cite{Louie,Stuttgart}. To
eventually resolve the issue, it is crucial to extract the bosonic
spectral function quantitatively so that a direct comparison between
the experiments and the calculations could be
possible~\cite{JunrenMEM,XJZhouMEM,NonBi2201}.

In this paper, we introduce a new approach which can unambiguously
determine the bare quasi-particle dispersion, and hence enables the
quantitative determination of the bosonic spectral function. The
method utilizes the subtle change of the quasi-particle dispersion
induced by the temperature. This is made possible by the super-high
resolution laser-based ARPES measurements, which present data of the
significantly higher quality feasible for a quantitative analysis of
a modified maximum entropy method~\cite{JunrenMEM}. By carrying out
such a procedure on a heavily over-doped
(Bi,Pb)$_2$Sr$_2$CuO$_{6+\delta}$ (Pb-Bi2201, T$_c$$\sim$5 K), we
are able to determine the bare band and extract the bosonic spectral
function quantitatively along the nodal direction. Our results
present a conclusive evidence that the electron-boson coupling is
strong (with a mass enhancement factor $\lambda\approx 0.42$) even
in the heavily over-doped cuprate compounds, and it originates from
the electron coupling to the multiple phonon modes.  Moreover, our
approach is general enough to be applicable for many other
materials.

\begin{figure}[tbp]
\begin{center}
\includegraphics[width=1.0\columnwidth,angle=0]{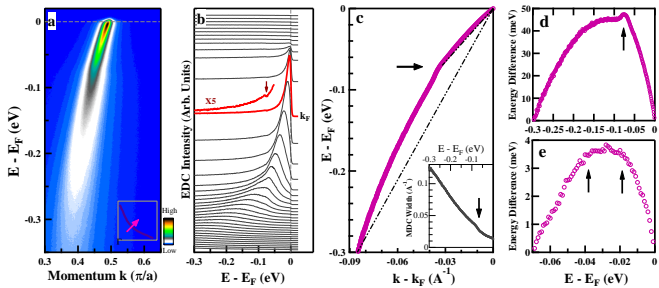}
\end{center}
\caption{Fine structure in the nodal dispersion of a heavily
over-doped Pb-Bi2201 ($T_c\sim 5$ K) measured at 15 K.  (a)
Photoemission image taken along the (0,0)-($\pi$,$\pi$) nodal
direction with the location of the momentum cut shown in the inset.
(b) Corresponding photoemission spectra (energy distribution curves,
EDCs). The EDC at the Fermi momentum $k_F$ (red curve) shows a dip
feature at $\sim 73$ meV, as indicated by an arrow in the expanded
spectrum. (c) Quasi-particle dispersion obtained by fitting MDCs
(momentum distribution curves) at different binding energies. The
inset shows the corresponding MDC width (FWHM). (d) Energy
difference between the measured dispersion and a straight line
connecting 0 ($E_F$) and $-0.3$~eV.  A peak at $\sim$73 meV is
clearly observed as indicated by an arrow. (e) Energy difference
between the measured dispersion and a straight line connecting 0 and
$-0.07$~eV. Two features can be identified at $\sim 41$ meV and
$\sim 17$ meV. }
\end{figure}

The angle-resolved photoemission measurements have been carried out
on our newly developed vacuum ultra-violet (VUV) laser-based ARPES
system~\cite{GDLiu}. The photon energy of the laser is 6.994 eV with
a bandwidth of 0.26 meV. The overall instrumental energy resolution
is set at 1 meV, which is significantly improved from 10$\sim$20 meV
of previous
measurements~\cite{PVBogdanov,ALanzara,PJohnson,AKaminski,XJZhouNature,AAKordyuk,WTZhang1,JunrenMEM,XJZhouMEM,NonBi2201}.
The angular resolution is $\sim 0.3$ degree, corresponding to a
momentum resolution $\sim$0.004 A$^{-1}$ at the photon energy of
6.994 eV, more than twice improved from 0.009 A$^{-1}$ at a usual
photon energy of 21.2 eV for the same angular resolution.  The
heavily overdoped (Bi,Pb)$_2$Sr$_2$CuO$_{6+\delta}$ (Pb-Bi2201)
single crystals with a $T_c\sim 5$~K were grown by the
traveling-solvent floating zone method. The samples are cleaved {\it
in situ} in vacuum with a base pressure better than 5 $\times$
10$^{-11}$ Torr.

\begin{figure}[bp]
\begin{center}
\includegraphics[width=1.00\columnwidth,angle=0]{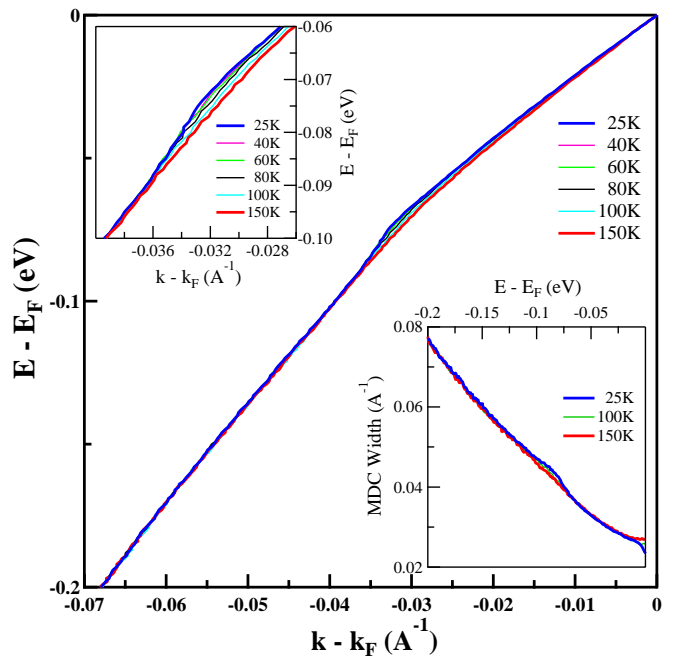}
\end{center}
\caption{ Temperature dependence of the nodal dispersion measured on
a heavily over-doped Pb-Bi2201 (T$_c$$\sim$5 K). The dispersions
near the kink region are expanded in the up-left inset. The
bottom-right inset shows MDC width measured at 25 K, 100 K and 150K.
}
\end{figure}

Figure~1(a) shows the raw data of photoemission image taken on a
heavily overdoped Pb-Bi2201 (T$_c$$\sim$5 K) sample along the
(0,0)-($\pi$,$\pi$) nodal direction at 15 K. Clear kink is observed
near 70 meV in the dispersion extracted from MDC (momentum
distribution curve) fitting (Fig.~1(c)), with a corresponding drop
in the MDC width (inset of Fig.~1(c)). In order to reveal the
possible fine structures in the dispersion, we plot the energy
difference between the measured dispersion and a featureless
straight line connecting 0 ($E_F$) and $-0.3$~eV on the dispersion
(Fig.~1(d)).  A prominent peak shows up near $\sim$73 meV. In
addition, in the energy difference between the measured dispersion
and another straight line connecting 0 and $-0.07$~eV (Fig. 1e), one
can also identify low energy features at $\sim 41$~meV and $\sim
17$~meV, signified by the slope changes.  We note that these three
features at $\sim$73meV, $\sim$41meV, and $\sim$17meV  are robust
and have been reproduced in several independent measurements.
Moreover, a clear dip at 73 meV (Fig. 1(b)) can be observed on the
EDCs (energy distribution curves) near the Fermi momentum ($k_F$).

The underlying bosonic spectral function $\alpha^2F(\omega)$ can be
extracted from the measured real part of the electron
self-energy~\cite{JunrenMEM}. The real part of self-energy is
related to $\alpha^2F(\omega)$ by:
\begin{equation}
 \mathrm{Re}\Sigma(\epsilon;T)= \int^{\infty}_0 \mathrm{d}\omega\,
 K \left(\frac{\epsilon}{kT},\frac{\omega}{kT} \right) \alpha^2F(\omega)\, ,
\end{equation}
where $K(y,y')=\int^{\infty}_{-\infty} dx\,f(x-y)2y'/(x^2-y'^2)$
with $f(x)$ being the Fermi distribution function~\cite{GGrimvall}.
In the ARPES measurements, the real part of the electron self-energy
is determined from the measured MDC dispersion $\epsilon_k$ by
Re$\Sigma(\epsilon_k)=\epsilon_k-\epsilon^0_{k}$, where $
\epsilon^0_k$ is the bare quasi-particle dispersion for an {\it
ad-hoc} system with the electron-boson coupling turned
off~\cite{BareBNote}. Attempts have been made to invert the
``effective" bosonic spectral function $\alpha^2F(\omega)$ from the
measured MDC dispersion by assuming an empirical bare band of a
quadratic form $\epsilon^0_k =
a(k-k_F)^{2}+b(k-k_F)^{2}$~\cite{JunrenMEM,XJZhouMEM,NonBi2201}. It
was found that the qualitative features of the bosonic spectral
function do not sensitively depend on the choice of the bare band.
However, the arbitrariness in choosing the appropriate parameters
for the bare band prevents a quantitative determination of the
bosonic spectral function~\cite {JunrenMEM,XJZhouMEM,NonBi2201}.
Here we propose a general approach to circumvent this problem by
making use of the self-energy change at different temperatures. We
assume that $\alpha^2F(\omega)$ and the bare band have negligible
temperature dependence over a relatively small temperature window,
and the Eliashberg function can be extracted from the self-energy
difference at different temperatures by:
\begin{multline}
 \mathrm{Re}\Sigma(\epsilon;T_1)-\mathrm{Re}\Sigma(\epsilon;T_2)=
 \int^\infty_0 \mathrm{d}\omega\, \alpha^2F(\omega) \\
 \times \left[K\left(\frac{\epsilon}{kT_1},\frac{\omega}{kT_1}\right)
 -K\left(\frac{\epsilon}{kT_2},\frac{\omega}{kT_2}\right)\right] \,
 \label{DSigma}
\end{multline}
The successful application of the approach relies critically on the
high quality of data, as it utilizes a small temperature-induced
change between two dispersions subjected to noises. Such a stringent
requirement on the data quality can now be met by the super-high
resolution laser-ARPES measurements(Fig. 2).  In this paper, we test
our new approach on the heavily overdoped Pb-Bi2201, in which the
electron-electron correlation is relatively weak, and
Eq.~(\ref{DSigma}) is believed to be applicable.

Figure 3 shows the detailed procedure to extract the bare band
dispersion and the bosonic spectral function.  To obtain the bosonic
spectral function $\alpha^2F(\omega)$ from Eq.~(\ref{DSigma}), one
needs to know the difference of the real part of self-energy
$\Delta\mathrm{Re}\Sigma(\epsilon) \equiv
\mathrm{Re}\Sigma(\epsilon;T_1)-\mathrm{Re}\Sigma(\epsilon;T_2)$ at
two different temperatures, as schematically shown in Fig. 3(a).
Here, we use two dispersions measured at 25 K and 100 K. Note that
$\Delta\mathrm{Re}\Sigma(\epsilon)$ depends on the selection of the
bare band dispersion.  As the result, an iteration approach is
needed, as detailed in the following: (1) Determine
$\Delta\mathrm{Re}\Sigma(\epsilon)$ based on the current selection
of the bare band dispersion.  The initial value
($\Delta\mathrm{Re}\Sigma_0(\epsilon)$) is taken as the direct
difference between the two dispersions, as schematically shown in
Fig. 3(a).  The result is shown in Fig.~3(c); (2) Extract the
``bosonic spectral function" SF from
$\Delta\mathrm{Re}\Sigma(\epsilon)$ by Eq.~(\ref{DSigma}), using a
maximum entropy method (MEM) similar to that in
Ref.~[\onlinecite{JunrenMEM}] but with the  modified integral
kernel. The result is shown in Fig.~3(d).  Note that the significant
change of the real part of the self-energy is limited to the energy
up to $\sim 100$~meV, above which the self-energy difference
fluctuates around zero and is most likely due to the data noise, as
shown in Fig.~3(c).  Therefore, we assume a maximum bosonic mode
energy 100~meV when doing the MEM analysis;   (3) Calculate the real
part of self-energy at 25K from the extracted bosonic spectral
function using Eq.~(1). The result is shown in Fig. 3(d); (4)
Determine the bare band dispersion by taking the difference between
the measured dispersion and the calculated real part of the
self-energy, both at 25K.  The result is shown in Fig.~3(b).  These
steps iterate until the result converges.  We find that the
procedure converges quickly and only three iterations are needed for
this analysis.

\begin{figure}[tbp]
\begin{center}
\includegraphics[width=0.96\columnwidth,angle=0]{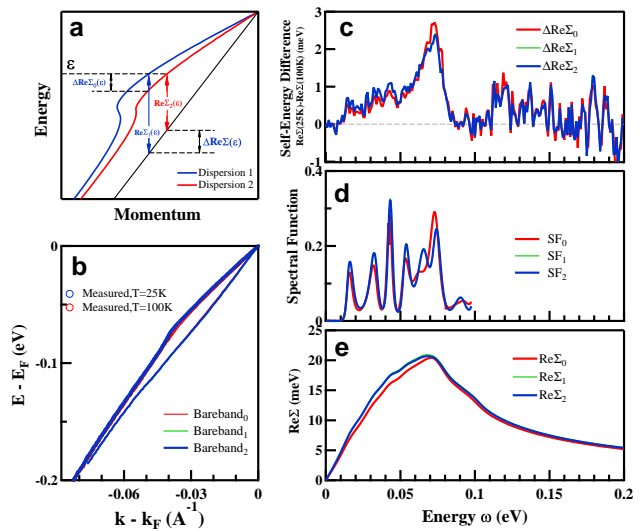}
\end{center}
\caption{Extraction procedure of the bare band and the Eliashberg
spectral function. (a) Schematic plot for the two dispersions and a
bare band to show the real parts of self-energies and their
difference for a given energy $\epsilon$.   (b) Measured nodal
dispersions of heavily overdoped Pb-Bi2201 at 25K and 100K and the
extracted bare bands for different iterations. (c) Self-energy
difference between the 25K and 100K dispersions for different
iterations. (d) Extracted Eliashberg spectral function for different
iterations. (e) Real part of electron self-energy calculated from
the extracted spectral function at 25 K for different iterations. }
\end{figure}

Fig. 4 summarizes the final results of the bare band (Fig. 4(a)),
Eliashberg spectral function (Fig. 4(b)), and electron self-energy
(inset of Fig. 4(a)). Although the results are obtained by using the
two representative dispersions at 25K and 100K,  the self-energy
calculated from the extracted Eliashberg function (Fig. 4(b)) at
other temperatures (here we show 60K data) matches very well with
the measured values, as shown in the inset of Fig. 4(a), indicating
the internal consistency of our approach. The obtained bare band is
close but less steeper than the one given by the LDA
calculations~\cite{NonBi2201PRB} (Fig. 4(a)), and is significantly
different from the empirical one used previously (green line in Fig.
4a)~\cite{NonBi2201}.

Sharp features are clearly identified in the extracted Eliashberg
spectral function (Fig. 4(b)). Five peaks at
(14-17),(30-32),(41-44),(54-58) and (70-74) meV can be identified
below 100 meV, and they are reproducible from several independent
measurements.  In fact, the features (14-17),(41-44) and (70-74) meV
are clearly visible in the raw data (Fig. 1).  These sharp features
show reasonably good agreement with the phonon modes observed in the
same material by infrared reflectance (Fig. 4(c))~\cite{IRBi2201}
and Raman scattering (Fig. 4(d))~\cite{RamanBi2201}. Note that these
optical measurements only see phonon modes with zero wavevector, and
the full band of phonons will exhibit dispersion in the Brillouin
zone, in particular for the higher energy phonon near 70$\sim$80
meV~\cite{Neutron}. Considering the rather weak magnetic
signal~\cite{MagneticOverdoped}, and the absence of magnetic mode at
15 K for the heavily overdoped Pb-Bi2201 sample (T$_c$$\sim$5 K),
our present results leave no doubt that the $\sim 70$ meV dispersion
kink is due to electron coupling with the multiple phonon modes.

\begin{figure}[tbp]
\begin{center}
\includegraphics[width=0.96\columnwidth,angle=0]{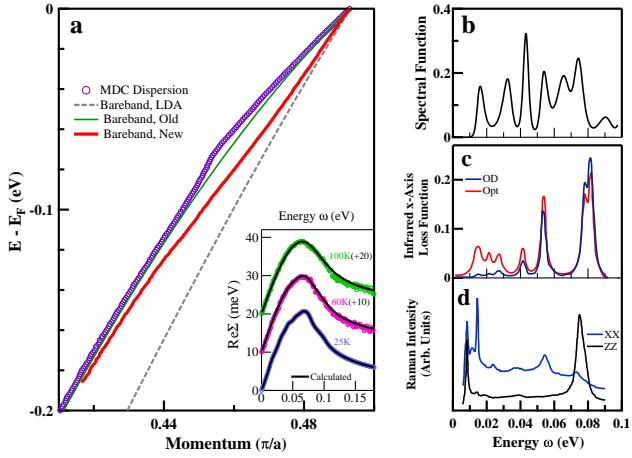}
\end{center}
\caption{Extracted bare band dispersion, self-energy and Eliashberg
spectral function of heavily over-doped Pb-Bi2201. (a) Measured
nodal dispersion (empty blue circles) at 25 K and the extracted bare
band (thick red line). For comparison, the bare band from the band
structure calculations (dashed grey line)~\cite{NonBi2201PRB} and
the bare band used by previous maximum entropy method (thin green
line)~\cite{NonBi2201PRB} are also shown. The bottom-right inset
shows the measured electron self-energy at different temperatures
and their comparison with those calculated from the extracted
Eliashberg spectral function. (b) Extracted Eliashberg functions
$\alpha^{2}F(\omega)$. (c) Loss function from infrared measurements
on optimally-doped (red line) and over-doped (blue line)
Bi-2201~\cite{IRBi2201}. (d) Raman spectra of Bi2201 under different
polarization geometries~\cite{RamanBi2201}.  }
\end{figure}

With the Eliashberg function extracted, it is possible to determine
quantitatively the strength of the electron-boson coupling. We
calculate the mass enhancement factor
$\lambda$=2$\int^\infty_0\frac{d\omega}{\omega}\alpha^2F(\omega)$,
and obtain $\lambda\approx 0.42$. As discussed above, the
contribution is mainly from the electron-phonon coupling. Our study
presents a conclusive evidence that the electron-phonon coupling is
strong even in the heavily over-doped cuprate samples. It is much
stronger than the value suggested by the first principles
calculations (0.14$\sim$0.20)~\cite{Louie}. Our present work thus
asks for a re-examination of theoretical calculations on the
strength of electron-phonon coupling in these compounds.

In summary, by taking advantage of the high quality data of the
laser-ARPES measurements, we have developed a new approach to
determine the bare band dispersion and extract Eliashberg spectral
function quantitatively. The method is general and can be applied to
many other materials, thus promoting ARPES technique to be
quantitative in probing many-body effects. Our results on the
heavily overdoped Pb-Bi2201 demonstrate unambiguously that the
$\sim$70 meV nodal dispersion kink is due to electron coupling to
the multiple phonons. In particular, we present a conclusive
evidence that the electron-phonon coupling is strong in cuprates
even in the heavily over-doped regime.


\begin{thebibliography}{99}
\bibitem{BareBNote} In a real system, the "bare band" is somewhat ambiguous because the
retarded effects of the electron-electron interaction could also act
as the effective bosonic modes. Therefore, the ``bare band''
actually depends on the energy scale we are looking at. For phonon
modes which usually have much smaller energy, the so called "bare
band" actually already includes the renormalization effects from the
other higher energy collective excitations.
\end{thebibliography}
\end{document}